\documentclass[aps,prl,twocolumn,superscriptaddress]{revtex4}
\usepackage{graphicx}
\usepackage{bm}

\begin{document}
\newcommand{\beq}{\begin{equation}}
\newcommand{\eeq}{\end{equation}}
\title{Second wind of the Dulong-Petit Law at a quantum critical point}
\author{V.~A.~Khodel}
\affiliation{Russian Research Center Kurchatov Institute, Moscow,
123182, Russia} \affiliation{McDonnell Center for the Space
Sciences \& Department of Physics, Washington University,
St.~Louis, MO 63130, USA}
\author{J.~W.~Clark}
\affiliation{McDonnell Center for the Space Sciences \& Department
of Physics, Washington University, St.~Louis, MO 63130, USA}
\author{V.~R.~ Shaginyan}\email{vrshag@thd.pnpi.spb.ru}
\affiliation{Petersburg Nuclear Physics Institute, RAS, Gatchina,
188300, Russia}
\author{M.~V.~Zverev}
\affiliation{Russian Research Center Kurchatov Institute, Moscow,
123182, Russia}\affiliation{ Moscow Institute of Physics and
Technology, Moscow, 123098, Russia }
\begin{abstract}
Renewed interest in $^3$He physics has been stimulated by
experimental observation of non-Fermi-liquid behavior of dense
$^3$He films at low temperatures. Abnormal behavior of the specific
heat $C(T)$ of two-dimensional liquid $^3$He is demonstrated in the
occurrence of a $T$-independent $\beta$ term in $C(T)$.  To uncover
the origin of this phenomenon, we have considered the group
velocity of transverse zero sound propagating in a strongly
correlated Fermi liquid. For the first time, it is shown that if
two-dimensional liquid $^3$He is located in the vicinity of the
quantum critical point associated with a divergent quasiparticle
effective mass, the group velocity depends strongly on temperature
and vanishes as $T$ is lowered toward zero.  The predicted vigorous
dependence of the group velocity can be detected in experimental
measurements on liquid $^3$He films.  We have demonstrated that the
contribution to the specific heat coming from the boson part of the
free energy due to the transverse zero-sound mode follows the
Dulong-Petit Law. In the case of two-dimensional liquid $^3$He, the
specific heat becomes independent of temperature at some
characteristic temperature of a few mK.
\end{abstract}

\maketitle

Almost two hundred years ago, Pierre-Louis Dulong and
Alexis-Th\'er\`ese Petit \cite{dulongpetit} discovered
experimentally that the specific heat $C(T)$ of a crystal is close
to constant independent of the temperature $T$.  This behavior,
attributed to lattice vibrations---i.e.\ phonons---is known as the
Dulong-Petit Law. Later, Ludwig Boltzmann \cite{boltzmann}
reproduced the results of Dulong and Petit quantitatively in terms
of the equipartition principle.  However, subsequent measurements
at low temperatures demonstrated that $C(T)$ drops rapidly as the
$T$ is lowered toward zero, in sharp contrast to Boltzmann's
theory.  In 1912, Peter Debye \cite{debye} developed a quantum
theory for evaluation of the phonon part of the specific heat of
solids, correctly explaining the empirical behavior $C(T) \sim T^3$
of the lattice component as $T \to 0$.  In the Debye theory, the
$T$-independence of $C(T)$ is recovered at $T \geq T_D$, where
$T_D$ is a critical temperature corresponding to the saturation of
the phonon spectrum.  With the advent of the Landau theory of
quantum liquids \cite{landau}, predicting a linear-in-$T$
dependence of $C(T)$ for the specific heat contributed by itinerant
fermions, our understanding of the low-temperature thermodynamic
properties of solids and liquids thus seemed to be firmly
established.  However, recent measurements
\cite{greywall,saunders2} of the specific heat of two-dimensional
(2D) $^3$He as realized $^3$He films absorbed on graphite preplated
with a $^4$He bilayer, reveal behavior strongly antithetical to
established wisdom, which calls for a new understanding of the
low-temperature thermodynamics of strongly correlated many-fermion
systems.

Owing to its status as a fundamental exemplar of the class of
strongly interacting many-fermion systems, liquid $^3$He remains a
valuable touchstone for low-temperature condensed-matter physics.
In recent years, interest in $^3$He physics has been driven by the
the observation of non-Fermi-liquid (NFL) behavior of dense $^3$He
films at the lowest temperatures $T\simeq 1$ mK reached
experimentally \cite{greywall,godfrin1,godfrin2,fukuyama,
morishita, saunders1,saunders2}.  In particular, measurements
of the specific heat $C(T)$ in the 2D $^3$He system show the
presence of a term $\beta$ tending to a finite value as
$T \to 0$.  Such behavior contrasts sharply with that of its
counterpart, three-dimensional (3D) liquid $^3$He; for this
system, the lower the temperature, the better Landau
Fermi-liquid (FL) theory works.  (Here we shall not consider
superfluid phases of $^3$He.)

\begin{figure} [! ht]
\begin{center}
\includegraphics [width=0.47\textwidth]{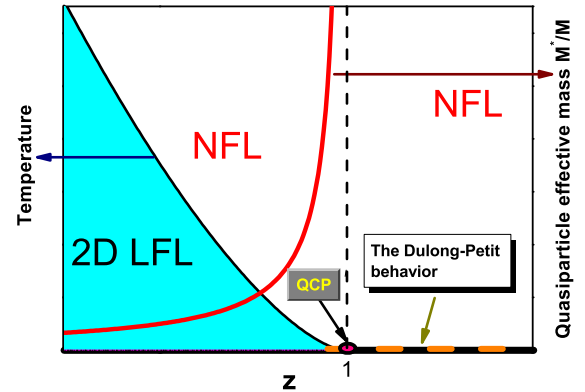}
\end{center}
\vspace*{-1.0cm} \caption{Phase diagram of the 2D liquid $\rm ^3He$
system.  The region defined by $z=\rho/\rho_{FC}<1$ is divided into
LFL and NFL domains separated by a solid line. The dependence
$M^*(z)\propto (1-z)^{-1}$ is shown by the solid line approaching
the dashed asymptote, thus depicting the divergence of the
effective mass at the quantum critical point $(z=1,T=0)$ indicated
by the arrow. In the region $z\gtrsim 1$, the fermion condensate
(FC) sets in and Dulong-Petit behavior of the specific heat is
realized for the strongly correlated quantum many-fermion system
(as represented by the dash horizontal line at $T=0$).}\label{fd}
\end{figure}

In seeking the origin of the anomalous contribution $\beta$
remaining in $C(T)$ at the lowest temperatures attained, it is
instructive to examine the schematic low-$T$ phase diagram of 2D
liquid $^3$He shown in Fig.~1.  The essential features of this
picture are documented by the cited experiments on $^3$He films.
The effective coupling parameter is represented by $z =
\rho/\rho_\infty$, where $\rho$ is the number density of the system
and $\rho_\infty$ is the critical density at which a quantum
critical point (QCP) occurs.  This QCP is associated with a
divergence of the effective mass $M^*(z)$, portrayed in Fig.~1 by
the curve (in red on line) that approaches the dashed asymptote at
$z=1$. The $T-z$ phase plane is divided into regions of 2D
Fermi-Liquid (FL) and non-Fermi-Liquid (NFL) behavior.  The part of
the diagram where $z<1$ consists of a FL region at lower $T$ and a
NFL region at higher $T$, separated by a solid curve.  The regime
where $z \gtrsim 1$ belongs to a NFL state with specific heat
taking a finite value $\beta(\rho)$ at very low temperatures. The
physical source of this excess heat capacity has not been
established with certainty, although it is supposed that the
$\beta$ anomaly is related to peculiarities of the substrate on
which the $^3$He film is placed.

In this letter we propose that the observed $\beta$ term in $C(T)$
can instead have its origin in an intrinsic mechanism analogous to
that producing the classical Dulong-Petit behavior in solids. It is
shown for the first time that in systems (such as 2D $^3$He)
containing a fermion condensate (FC), the group velocity of
transverse zero sound depends strongly on temperature. It is this
dependence that gives rise to the $\beta$ term, granting the
Dulong-Petit Law a ``second wind.''

As indicated above, the most challenging feature of the NFL
behavior of liquid $^3$He films involves the specific heat $C(T)$.
According to Landau theory, $C(T)$ varies linearly with $T$, and at
low film densities the experimental behavior of the specific heat
of 2D liquid $^3$He is in agreement with FL theory. However, for
relatively dense $^3$He films, this agreement is found to hold only
at sufficiently {\it high} temperatures. If $T$ is lowered into the
millikelvin region, the function $C(T)$ ceases to fall toward zero
and becomes flat \cite{greywall,saunders1,saunders2}.

The common explanation \cite{greywall,saunders2,golov} of the
flattening of $C(T)$ seen in these experiments imputes the
phenomenon to disorder associated with the substrate that supports
the $^3$He film.  More specifically, it is considered that there
exists weak heterogeneity of the substrate (steps and edges on its
surface) such that quasiparticles, being delocalized from it, give
rise to the low-temperature feature $\beta$ of the heat capacity
\cite{saunders2}. Even if we disregard certain unjustified
assumptions made in Ref.~\onlinecite{golov}, there remains the
disparate fact that the emergent constant term in $C(T)$ is of
comparable order for different substrates
\cite{saunders1,saunders2,greywall}.  Furthermore, the explanation
posed in Ref.~\cite{golov} implies that the departure of $C(T)$
from FL predictions shrinks as the film density increases, since
effects of disorder are most pronounced in weakly interacting
systems. Contrariwise, the anomaly in $C(T)$ makes its appearance
in the density region where the effective mass $M^*$ is greatly
enhanced \cite{saunders1,saunders2} and the 2D liquid $^3$He system
becomes strongly correlated. This reasoning compels us to consider
that the NFL behavior of $C(T)$ is an {\it intrinsic} feature of 2D
liquid $^3$He, which is associated with the divergence of the
effective  mass rather than with disorder. The flattening of the
curve $C(T)$ as seen in $^3$He films is by no means a unique
phenomenon.  Indeed, as expressed in the Dulong-Petit (DP) law, the
specific heat $C(T)$ of solids remains independent of $T$ as long
as $T$ exceeds the Debye temperature $\Omega_D$, which is
determined by the saturation of the phonon spectrum of the crystal
lattice.  Normally, the value of $\Omega_D$ is sufficiently high
that the DP law belongs to classical physics. However, we will
argue that the DP behavior of $C(T)$ can also make its appearance
at extremely low temperatures in strongly correlated Fermi systems,
with zero sound playing the role of phonons.

To clarify the details of this phenomenon and calculate the
specific heat $C(T)$, we evaluate a part $F_B$ of the free energy
$F$ associated with the collective spectrum $\omega(k)=ck$, based
on the standard formula \beq F_B =\int\limits_0^{\infty}
{d\omega\over\pi}{1\over e^{{\omega/T}}-1}\int{\rm Im} \left[\ln
D^{-1}(k,\omega)\right] d\upsilon \label{fre}, \eeq where
$D(k,\omega)$ is the boson propagator, and $d\upsilon$ is an
element of momentum space.  Upon integration by parts this formula
is recast to \beq F_B=T\int\limits_0^{\infty} {d\omega\over\pi} \ln
\left(1-e^{-{\omega/T}}\right) \int {\rm Im}\left({{\partial
D^{-1}(k,\omega)/\partial\omega}\over
D^{-1}(k,\omega)}\right)d\upsilon. \eeq If damping of the
collective branch is negligible (the case addressed here), then
$D^{-1}(k,\omega)\simeq(\omega-ck)$ and $\partial
D^{-1}(k,\omega)/\partial\omega \simeq 1$, while ${\rm
Im}D^{-1}(k,\omega)\simeq \delta(\omega-ck)$, and we arrive at the
textbook formula \beq F_B=T\int \ln \left( 1-e^{-{ck/ T}}\right)
\theta(\Omega_0-ck) d\upsilon, \label{textf} \eeq where $\Omega_0$
is the characteristic frequency of zero sound. At $T\gg \Omega_0$,
the factor $\ln (1-e^{-{\omega/T}})$ reduces to $\ln (\omega/T)$,
yielding the result \beq F_B(T)\propto T\ln (\Omega_0/T), \eeq
which, upon the double differentiation, leads to the DP law $C(T)=
{\rm const.}$ At first sight, this law has nothing to do with the
situation in 2D liquid  $^3$He.  Its Fermi energy $\epsilon^0_F$ is
around 1 K at densities where the Sommerfeld ratio $C(T)/T$ soars
upward as $T\to 0$, while $\Omega_0$ must be lower than $T\simeq 1$
mK.  Indeed, in any conventional Fermi liquid, including 3D liquid
$^3$He, there is no collective degree of freedom whose spectrum is
saturated at such low ratios $\Omega_0/\epsilon^0_F$.

This conclusion remains valid {\it assuming} 2D liquid $^3$He is an
ordinary Fermi liquid.  However, as seen from Fig.~\ref{fd}, if the
quantum critical point is reached at $T\to0$ and some critical
density $\rho_{\infty}$ where the effective mass $M^*(\rho)$
diverges, as it does in the present case
\cite{godfrin1,godfrin2,saunders1,saunders2,prl100}, the situation
changes dramatically. This is demonstrated explicitly in the
results of standard FL calculations of the velocity $c_t$ of
transverse zero sound, which satisfies \cite{halat,halat1} \beq
{c_t\over 2v_F}\ln {c_t+v_F\over c_t-v_F}-1= {F_1-6\over
3F_1(c^2_t/v^2_F-1)}, \eeq where $v_F=p_F/M^*$ is the Fermi
velocity and $F_1=p_FM^*f_1/\pi^2$ is a dimensionless version of
the Landau first harmonic $f_1$ \cite{pfit,pfit1,halat}.   The
divergence of the effective mass $M^*$ at the QCP implies that at
the critical density determined by $f_1p_FM/\pi^2=3$
\cite{pfit,pfit1}, one has \beq c^2_t(\rho)\simeq{p_F^2\over
5M^*(\rho)M} \to0, \label{clim} \eeq whereas the sound velocity
$c_s$ remains finite in this limit \cite{halat,halat1,nozz}.

We see then that in case the effective mass $M^*$ diverges, the
group velocity $c_t$ vanishes as $\sqrt{1/M^*}$.  Flattening of
the single-particle spectrum $\epsilon(p)$ prevails as long as
$|p-p_F|/p_F<M/M^*$, implying that the transverse mode softens
only for rather small wave numbers $k\sim p_FM/M^*$.
Unfortunately, the associated numerical prefactor cannot be
established, rendering estimation of $\Omega_0\sim
(p^2_F/M)\sqrt{M/M^*}$ correspondingly uncertain.
Nevertheless, one cannot exclude a significant enhancement
of the Sommerfeld ratio $C(T)/T$ at $T\simeq 1$ mK due to
softening of the transverse zero sound in the precritical density
region.

At $T\to0$ and densities exceeding $\rho_\infty$, the system
undergoes a cascade of topological phase transitions in which the
Fermi surface acquires additional sheets \cite{zb,shagp,prb2008}.
As indicated in Fig.~\ref{fd}, FL theory continues to hold with
quasiparticle momentum distribution $n(p)$ satisfying $n^2=n$,
until a greater critical density $\rho_{FC}$ is reached where a new
phase transition, known as fermion condensation, takes place
\cite{ks,vol,noz,physrep,shagrev,prb2008,shagrev1}. Beyond the
point of fermion condensation, the single-particle spectrum
$\epsilon(p)$ acquires a flat portion. The range $L$ of momentum
space adjacent to the Fermi surface where the FC resides depends on
the difference between the effective coupling constant and its
critical value. As will be seen, $L$ is a new dimensional parameter
that serves to determine the key quantity $\Omega_0$.

At finite $T$, the dispersion of the FC spectrum $\epsilon(p)$
existing at $\rho>\rho_{FC}$ acquires a nonzero value proportional
to temperature \cite{noz,physrep,shagrev,prb2008}: \beq
\epsilon(p,T)=T\ln{1-n_*(p)\over n_*(p)},\quad p_i<p<p_f,
\label{spt} \eeq where $0<n_*(p)<1$ is the FC momentum distribution
and $p_i$ and $p_f$ are the lower and upper boundaries of the FC
domain in momentum space. Consequently, in the whole FC region, the
FC group velocity, given by \beq v(p,T)={\partial\epsilon(p)\over
\partial p}=-T  {\partial n_*(p)/\partial p\over
n_*(p)(1-n_*(p))},\quad p_i<p<p_f, \label{vfc} \eeq is proportional
to $T$. Significantly, in the density interval
$\rho_{\infty}<\rho<\rho_{FC}$ the formula (\ref{spt}) describes
correctly the single-particle spectrum $\epsilon(p,T)$ in case the
temperature $T$ exceeds a very low transition temperature
\cite{prb2008}. The FC itself contributes a $T$-independent term to
the entropy $S$; hence its contribution to the specific heat
$C(T)=T dS/dT$ is zero.  Accordingly, we focus on the zero-sound
contribution to $C(T)$ in systems having a FC.

Due to the fundamental difference between the FC single-particle
spectrum and that of the remainder of the Fermi liquid, a system
having FC is, in fact, a two-component system.  Remarkably, the FC
subsystem possesses its own set of zero-sound modes, whose wave
numbers are relatively small, not exceeding $L=(p_f-p_i)>0$.  The
mode of prime interest for our analysis is that of transverse zero
sound.  As may be seen by comparison of formulas (\ref{clim}) and
(\ref{vfc}), its velocity $c_t$ {\it depends on temperature} so as
to vanish like $\sqrt{T}$ as $T \to 0$.

To verify the latter property explicitly, we observe first that for
systems with a rather small proportion of FC, evaluation of the spectrum of
collective excitations may be performed by employing the familiar FL
kinetic equation \cite{trio,halat}
\beq
\left(\omega-{\bf k}{\bf v}\right)\delta n({\bf  p})
=-{\bf k}{\bf n}{\partial n(p)\over \partial p}
\int {\cal F}({\bf p},{\bf p}_1) \delta n({\bf p}_1)d\upsilon_1.
\label{kin}
\eeq
Focusing on transverse zero sound in 2D liquid $^3$He, one need
only retain the term in the Landau amplitude ${\cal F}$ proportional
to the first harmonic $f_1$.  To proceed further we make the
usual identification
$(c_t-\cos\theta)\delta n({\bf p})=\left(\partial n(p)/\partial
p\right)\phi({\bf n})$, where $\cos\theta={\bf k}{\bf v}/kv$.
Equation (\ref{kin}) then becomes
\beq
\phi(\theta) = -f_1p_F\cos\theta\int \cos\chi
{\partial n(p_1)/\partial p_1\over c_t-v(T)\cos\theta_1}
\phi(\theta_1) {dp_1d\theta_1\over (2\pi)^2},
\label{kins}
\eeq
where $\cos\chi= \cos\theta\cos\theta_1+\sin\theta\sin\theta_1$,
while $v(T) $ is given by equation (\ref{vfc}).  The solution describing
transverse zero sound is $\phi({\bf n})\sim \sin\theta\cos\theta$.

We see immediately that $c_t\gg v(T)\sim T$; therefore the
transverse sound in question does not suffer Landau damping. In
this situation, we are led to the simple result \beq
c^2_t=-{p_F\over 5M} \int {\partial n(p)\over \partial p} v(p,T)dp
\eeq upon keeping just the leading relevant term
$v(T)\cos\theta/c^2_t$ of the expansion of
$(c_t-v(T)\cos\theta)^{-1}$ and executing straightforward
manipulations.  Factoring out an average value of the group
velocity $v(p,T)\propto T/p_F$, we arrive at the stated behavior
\beq c_t(k) \simeq \sqrt{{T\over M}} \label{trs} \eeq for wave
numbers $k$ not exceeding the FC range $L$. Transverse sound can of
course propagate in the other, noncondensed subsystem of 2D liquid
$^3$He consisting of quasiparticles with normal dispersion
\cite{halat,halat1,trio}. However, its group velocity is
$T$-independent, so the corresponding contribution to the free
energy is irrelevant.

As noted above, the characteristic wave number of the soft
transverse zero-sound mode is given by the FC range
$L(\rho)=p_f-p_i$, treated here as an input parameter. The key
quantity $\Omega_0$ is therefore estimated as $\Omega_0\simeq
k_{\rm max}c_t$, where $k_{\rm max}$ is the maximum value of the
zero sound momentum at which the zero sound still exists. In our
case the zero sound is associated directly with the FC; hence
$k_{\rm max}\simeq \sqrt{Lp_F}$ and we have \beq \Omega_0\simeq
k_{\rm max}c_t\simeq \sqrt{{TLp_F\over M}}. \eeq As long as the
inequality $Lp_F/M<T$ is met (or equivalently, $T/\epsilon^0_F
>L/p_F$ holds), the ratio $\Omega_0/T$ is small, and we are led to
the DP result $C(T)={\rm const}$.  Then, in spite of the low
temperature, $C$ behaves as if the system were situated in the
classical limit rather than at the QCP. Such a behavior is ensured
by the fact that the system contains a macroscopic subsystem with
heavy quasiparticles.  As the temperature ultimately goes down to
zero at the fixed density $\rho$, the inequality $Lp_F/M<T$
eventually fails, the quantum regime is restored and the dominant
contribution to $C$ comes from the ``normal'' fermions. In other
words, there exists an extremely low temperature $T_0$ below which
the usual FL behavior of zero sound is recovered.

Interestingly, the value of the constant term in $C(T)$ can be
evaluated in closed form in terms of the FC range $L$.  Upon
inserting $\omega_t(k)=c_tk$ into Eq.~(\ref{textf}) and
integrating, the $T$-independent term in the specific heat is found
to be \beq {C\over N}\simeq{Lp_F\over 8\pi\rho} \label{crel} \eeq
where $N$ is the number of atoms in the film. The FC range
parameter $L$ also enters the result derived analogously for the
spin susceptibility $\chi$. The FC component of $\chi$ is given by
\cite{yak,prb2008} \beq \chi_*(T)\simeq\chi_C(T) {L\over p_F},
\label{chifc} \eeq where $\chi_C(T)=\mu^2_B\rho/T$.

The results (\ref{crel}) and (\ref{chifc}) jointly establish an
unambiguous relation within our model between the $T$-independent
term in the specific heat $C(T)$ and the Curie component of the
spin susceptibility $\chi(T)$ (which has also been observed
experimentally \cite{godfrin1,godfrin2}).  This relation can be
tested using existing experimental data \cite{saunders2}. The
$T$-independent specific heat $C/N$ exists in the density region
around $\rho=9.5$ nm$^{-2}$.  Being referred to one particle, it is
readily evaluated from $\beta\simeq0.25$ mJ/K.  One finds
$C/N\simeq0.01$, yielding $L/p_F\simeq 0.05$.  On the other hand,
the data for the spin susceptibility given in Fig.~2(B) of
Ref.~\onlinecite{saunders2} supports a Curie-like component at
$\rho=9.25$ nm$^{-2}$.  The value of the corresponding numerical
factor extracted from the data, which according to
Eq.~(\ref{chifc}) is to be identified with the ratio $L/p_F$, is
approximately 0.07. Given the uncertainties involved, our model is
consistent with the experimental data reported in \cite{saunders2}.

In summary, we have analyzed the group velocity of transverse zero
sound propagating in a strongly correlated Fermi liquid.  We have
shown for the first time that if two-dimensional liquid $^3$He is
located in the vicinity of the quantum critical point associated
with a divergent quasiparticle effective mass, the group velocity
depends strongly on temperature and vanishes at diminishing
temperatures. Such a vivid dependence of the group velocity can be
detected in experimental measurements on the liquid. We have
demonstrated that the contribution to the specific heat coming from
the boson part of the free energy contributed by the transverse
zero-sound mode follows the Dulong-Petit law. Accordingly, the
specific heat becomes independent of temperature at some
characteristic temperature. In the case of two-dimensional liquid
$^3$He, this temperature can be a few mK. At sufficiently lower
temperature the usual FL behavior of zero sound is recovered. The
model developed from the analysis is found to be in reasonable
agreement with experimental measurements.

This research was supported by the McDonnell Center for the Space
Sciences, by Grant No.~2.1.1/4540 from the Russian Ministry of
Education and Science, and by Grants \# 09-02-01284 and 09-02-00056
from the Russian Foundation for Basic Research.

\end{document}